\documentclass[12pt]{article}
\def\beqar {\begin{eqnarray}}
\def\eeqar {\end{eqnarray}}
\def\beq {\begin{equation}}
\def\eeq {\end{equation}}
\def\ra {{\rangle}}
\def\la {{\langle}}
\def\half {{\textstyle{1\over 2}}}
\def\Tr {{\rm Tr}}

\def\del {{\partial}}

\def\ep {{\epsilon}}
\def\vf {{\varphi}}

\def\bz {\bar{z}}

\def \vx {\vec{x}}

\def \C {{\cal C}}

\def \A {{\cal A}}

\def \P {{\cal P}}

\begin{document}

\begin{titlepage}
\null\vspace{-62pt}

\pagestyle{empty}
\begin{center}
\rightline{CCNY-HEP-03/2}

\vspace{1.0truein}
{\large\bf Some topological issues for ferromagnets}\\
\vspace {.1in}
{\large \bf and fluids}
\vskip .1in
\vspace{.5in}V.P. NAIR\\
\vspace{.1in}{\it Physics Department\\
City College of the CUNY\\
New York, NY 10031}\\
E-mail: vpn@sci.ccny.cuny.edu\\
\vspace {.3in}
RASHMI RAY\\
\vskip .1in{\it Physical Review Letters\\
American Physical Society\\
Ridge, NY 11367}\\
E-mail: ray@aps.org\\
\vspace{0.1in}
\end{center}
\vspace{0.5in}

\centerline{\bf Abstract}

We analyze the canonical structure of a continuum model of
ferromagnets and clarify known difficulties
in defining a momentum density.
The moments of the momentum
density corresponding to volume-preserving
coordinate transformations can be defined, but
a nonsingular definition of the other moments requires
an enlargement of the phase space which illuminates a close relation
to fluid mechanics.
We also discuss the nontrivial connectivity of the phase space
for two and three dimensions
and show how this feature can be
incorporated in the quantum theory,
working out the two-dimensional case in some
detail.

\end{titlepage}

\hoffset=0in
\newpage
\pagestyle{plain}
\baselineskip=16pt
\setcounter{page}{2}
\newpage

\section{Introduction}

It has been known for some time that there is a problem
in defining a suitable momentum density for
ferromagnets,
as noted in \cite{haldane, bishop}.
The difficulty has to do with the global topology of the
space of fields \cite{haldane}.
One can certainly define an adequate
momentum density
in the spin wave approximation, and even express
it in terms of globally
defined variables. But such an expression does not
preserve the rotational symmetry of the problem \cite{bishop}.
A nonsingular expression which preserves the rotational symmetry is
not possible. In this paper, we shall consider the continuum model of
ferromagnets and analyze the canonical structure of the theory.

We show how these results can be understood
and clarified by standard canonical
quantization of the
continuum action for the ferromagnet.
We show that it is possible to define the moments of the momentum
density corresponding to volume-preserving
coordinate transformations, but the other moments
are not unique and have singularities.
Our resolution of the singularities is based on enlarging the phase
space. We note that the phase space for the ferromagnet
is obtained from the standard
phase space for fluid mechanics by restricting
to the incompressible case. In other words, as far as the canonical
structure is concerned, the
ferromagnet may be viewed as an incompressible fluid (with a Hamiltonian
which is quite different from the fluid Hamiltonian).

Since volume-preserving moments include the total momentum and
total angular momentum, we may ask why we should need
a momentum density, why we need to address the singularities.
One simple reason has to do with excitations which can be locally
generated, say, by applying a focused beam of radiation to
a small region in the ferromagnetic lattice. Because of the localized
nature, it is the conservation law expressed in terms of
the divergence of
the momentum density  which is relevant when
the radiation is absorbed and magnetic excitations are generated.
This example shows that
our analysis is more than just formalistic
improvement.

We also discuss the connectivity of the phase space
for two and three dimensions. Generally, nontrivial
connectivity can lead to inequivalent quantum theories
as, for example, in the case of $\theta$-vacua in Chromodynamics.
The present situation is however more complicated, since
the nontrivial connectivity involves all the phase space
variables, which do not commute among themselves in the quantum theory.
By working out the two-dimensional case in some detail, we show
how one can incorporate this feature in the quantum theory.
Ferromagnetic systems with Skyrmion excitations,
the quantum Hall ferromagnet being a specific example,
can provide a physical context for our considerations \cite{sondhi}.
The $\theta$-term determines the
statistics of the Skyrme solitons. A full quantum treatment of these
excitations will require our formalism, since the noncommutativity of the
phase space variables involved in the definition of the $\theta$-term
must be taken into account.

In section 2, we discuss the canonical structure
of the ferromagnet. Section 3 is devoted to a similar analysis
for fluids and its relation to the ferromagnet.
General topological observations
are made in
section 4.
The quantization
of the two-dimensional case including the effects of the nontrivial
connectivity of the phase space is done in section 5.

\section{The canonical structure of the ferromagnet}

The degrees of freedom for the ferromagnet are the spin variables
$S_a(I)$ obeying the commutation rules
\beq
[ S_a (I), S_b (J)] = i \ep_{abc} S_c (I) ~\delta_{IJ}
\label{1}
\eeq
Here $a, b, c= 1, 2, 3$ and
the indices $I, J$ label the lattice sites.
The action for the ferromagnet should be such that canonical
quantization leads to the commutation rules given above.
For a single spin, this is well known.
The action can be taken as
\beq
{\cal S} = -i {n } \int dt~ \Tr (\sigma_3 g^{-1} {\dot g}) ~-~\int dt~ H
\label{2}
\eeq
Here $g$ is an element of the group $SU(2)$. It
may be taken to be
of the form $\exp ( it_a \theta^a )$ where $t_a = \half
\sigma_a$ and $\sigma_a$  are the Pauli
matrices.
The action (\ref{2}) is invariant under
$g\rightarrow g \exp ({\textstyle i\over
2}\sigma_3 \vf )$ modulo surface terms.
Therefore, the theory is defined on the
two-dimensional sphere $SU(2)/U(1) = S^2$.
The spin variables are given by
$\Tr (g \sigma_3 g^{-1} t_a)$ which can be shown to obey
the commutation rules (\ref{1}) upon
quantization.
$n$ will turn out to be an
integer with $\half n$ giving the maximal spin or $j$-value of the
spin variables. We shall consider spin-$\half$ variables
from now on, so that $n=1$ for all equations below.
The results can be trivially extended to arbitrary values of $n$.

For several independent spins, we may generalize this action
as
\beq
{\cal S}= -i \sum_I \int dt~ \Tr (\sigma_3 g_I^{-1} {\dot g}_I) ~-~
\int dt~ H
\label{3}
\eeq
The naive continuum limit of this action is thus given by
\beq
{\cal S} = -i \int {d^3x dt \over v }~ \Tr (\sigma_3 g^{-1} {\dot g})
~-~ \int dt~ H (S)
\label{4}
\eeq
$v$ is the spatial volume corresponding to a single spin,
determined by the unit cell of the lattice.
Notice that this action (\ref{4}) is invariant under $g\rightarrow g \exp
({\textstyle i\over 2}\sigma_3 \vf )$ modulo surface terms, where $\vf$
is now a function of $x$, so that the dynamical variables
are in $SU(2)/U(1)= S^2$. Since this is a
first order action,
$g$ will denote the phase space coordinates and with the above
observation, we can identify the phase space ${\cal P}$ as
\beq
{\cal P} = \left\{ {\rm set ~of ~ all ~ maps~}
{\bf R^3} \rightarrow SU(2)/U(1) \right\}\label{4a}
\eeq

We now consider the canonical quantization of
this action. The surface term in $\delta {\cal S}$ resulting from the
partial integration over the time-variable $t$ is
\beq
{\cal A} = -i \int {d^3x\over v}~\Tr ( \sigma_3 g^{-1} \delta g)
\label{5}
\eeq
This is the canonical or symplectic potential of the theory. Explicitly,
if $\theta^a$ are the group parameters
\beq
g^{-1}\delta g \equiv   -it_a \left( {\tilde E}\right) ^a_i \delta \theta^i,
\hskip .5in
\delta g g^{-1}\equiv  -it_a E^a_i \delta \theta^i
\label{6}
\eeq
and
\beqar
{\cal A} &=& - \int {d^3x \over v}~~{\tilde E}^3_i ~\delta \theta^i \equiv
\int {\cal A}_i (x) ~\delta \theta^i (x)\nonumber\\
&\equiv& {\cal A}_I ~\delta\theta^I\label{7}
\eeqar
where we have introduced a composite index $I=(i,x)$ to avoid clutter
in notation.

The standard formula
\beq
\delta e^{\hat{A}} = \int^{1}_{0} d\gamma ~e^{\gamma \hat{A}} ~\delta
\hat{A} ~e^{(1-\gamma ) \hat{A}}
\eeq
may be used to determine ${\tilde E}^{a}_{i}$ and $E^{a}_{i}$ in terms of the group parameters.

It may be useful at this stage to recall some general features
of the canonical formalism \cite{canonical}. Consider a general
symplectic potential
\beq
\A = \A_I ~\delta \xi^I = \int d^3x~ \A_i (x) ~\delta \xi^i (x)
\label{6a}
\eeq
The
functional curl of
${\cal A}$ gives the canonical two-form or symplectic structure
$\Omega_{ij} (x,x')$ whose inverse will give the fundamental
Poisson brackets.
Explicitly
\beqar
\Omega_{IJ}&=& \delta_I \A_J - \delta_J \A_I\nonumber\\
\Omega_{ij}(x, x') &=&{\delta \over \delta \xi^i( x )}\A_j(x '
) ~-~ {\delta \over \delta \xi^j(x ')} \A_i (x )
\label{7a}
\eeqar
where we have again used the composite index notation
$I= (i, x )$ and $J=(j, x')$ and
$\delta_I ={\delta /\delta\xi^i(x )}$.

With the understanding that the variations are
antisymmetrized, we may write $\Omega$ as a differential two-form
\beqar
\Omega &=& \delta {\cal A}
={1\over 2} \Omega_{IJ}~\delta\xi^I \wedge \delta \xi^J
\nonumber\\
&=& {1\over 2} \int \Omega_{ij}(x,x') ~\delta \xi^i (x)
\wedge\delta \xi^j (x')
\label{8}
\eeqar
Notice that from the definition of $\Omega$, we have the identity
\begin{equation}
\delta_I \Omega_{JK} + \delta_J \Omega_{KI} + \delta_K \Omega_{IJ}
=0\label{9a}
\end{equation}

We now turn to
canonical transformations and their generators.
Let
$\xi^I \rightarrow \xi^I +a^I(\xi)$ be an infinitesimal
transformation
of the canonical variables. This transformation is canonical if it
preserves the canonical structure $\Omega$. The change in $\Omega$
arises from two sources, due to the $\xi$-dependence of the components
$\Omega_{IJ}$ and due to the fact that $\Omega_{IJ}$ transforms under
change of phase space coordinate frames. (The latter is the change due to
the frame factors $\delta \xi^I \wedge \delta \xi^J$
in writing $\Omega$ as a differential form.)
The total change is
\beqar
\Delta \Omega_{IJ} &=& \delta_I \alpha_J - \delta_J \alpha_I\nonumber\\
\alpha_I &=& a^K\Omega_{KI}\label{9b}
\eeqar
where we have used (\ref{9a}). (This change is also the so-called
Lie derivative of $\Omega$.)
More explicitly, expanding out the composite notation,
\beqar
\Delta \Omega_{ij}(x, x ')&=& \left[ {\delta \over \delta
\xi^i(x) }\alpha_j(x') -{\delta \over \delta
\xi^j(x')}\alpha_i(x)\right]\nonumber\\
\alpha_i( x)&=&\int  d^3x'~ a^k(x')~\Omega_{ki}(x', x)
\label{9c}
\eeqar
The transformation $\xi^I \rightarrow \xi^I +a^I(\xi )$
may be represented by a vector field
$V = a^I \delta_I = \int_x a^i (x) (\delta / \delta \xi^i (x) )$.
The expression $\alpha_I = a^K \Omega_{KI}$ is then a one-form
or covariant vector corresponding to the contraction of
$V$ with $\Omega$, often denoted by $V \rfloor \Omega$.
In other words
\beqar
V\rfloor \Omega &\equiv& a^K \Omega_{KI} ~\delta \xi^I\nonumber\\
V&=& a^I {\delta \over \delta \xi^I}\label{9c1}
\eeqar
The change in $\Omega$ is thus the curl of $V\rfloor \Omega$,
which is denoted by
$ \delta (V \rfloor \Omega )$ in the language of differential forms.

Equation (\ref{9b}) shows that
the transformation $\xi^I\rightarrow \xi^I+a^I$ will preserve
$\Omega$, and hence be a canonical transformation, if
\beq
\delta_I \alpha_J - \delta_J \alpha_I =0\label{9b1}
\eeq
This condition may be solved
as
\begin{equation}
\alpha_I = -\delta_I ~G\label{9d}
\end{equation}
for some function $G$ of the phase space variables.
The function $G$ defined by this equation is
the generator of the canonical transformation.
( Equation (\ref{9b1}) is the general requirement for
a transformation to be canonical.
Equation (\ref{9d}) is a
necessary and sufficient condition locally on the phase space.
If the phase space has nontrivial topology,
the vanishing of $\Delta \Omega$ may have more general solutions.
Even though locally all solutions look like (\ref{9d}), the
$G$ so-defined may not
exist globally on the phase space. We will return to this question in
more detail later.)

The inverse of $\Omega$ is defined by $(\Omega^{-1})^{IJ} \Omega_{JK}
=\delta^I_K$ which expands out as
\beq
\int_V~d^3x' (\Omega^{-1})^{ij}(x, x') ~\Omega_{jk}(x', x'')
=\delta^i_k ~\delta^{(3)} (x-x'')\label{9e}
\eeq
Using the inverse
of $\Omega$, the Poisson bracket of functions $F, G$ on
phase space is defined as
\beqar
\{ F, G\}&=& (\Omega^{-1})^{IJ} \delta_I F \delta_J G\nonumber\\
&=&\int d^3x ~d^3x' ~ (\Omega^{-1})^{ij}(x, x ') {\delta
F\over \delta \xi^i(x )} {\delta G \over \delta \xi^j(x')}
\label{9f}
\eeqar
We may also rewrite equation (\ref{9d}) as
$a^I = (\Omega^{-1})^{IJ} \delta_J G$.
This shows that the change of any function $F$ under a canonical
transformation generated by $G$ is given by
\beqar
\delta F &=& a^I \delta_I F = (\Omega^{-1})^{IJ} \delta_J G ~\delta_I
F\nonumber\\
&=& \{ F, G\} \label{9g}
\eeqar
The change of any variable, so long as
the transformation is canonical, is given by the Poisson bracket
of the variable with the generating function for the transformation.

We will now apply these results to the ferromagnet.
With the understanding that the variations are
antisymmetrized, as emphasized by the wedge symbol,the symplectic form
$\Omega$ for the ferromagnet can be written as
the differential two-form
\beqar
\Omega &=& \delta {\cal A} = i \int {d^3x \over v} \Tr (\sigma_3
g^{-1}\delta g \wedge g^{-1}\delta g) \nonumber\\
&=& {1\over 2} \int \Omega_{ij}(x,x') ~\delta \theta^i (x)
\wedge\delta \theta^j (x')
\nonumber\\
&=& {1\over 2} \int {d^3x \over v} ~ \ep_{ab3} ~{\tilde E}^a_i ~{\tilde
E}^b_j ~\delta \theta^i \wedge \delta \theta^j\nonumber\\
\Omega_{ij}(x,x')&=& {1\over v}~\ep_{ab3} ~{\tilde E}^a_i~ {\tilde
E}^b_j ~\delta^{(3)}(x-x')\label{8}
\eeqar
The potential ${\cal A}$ is not invariant under $g \rightarrow g \exp( i
t_3 \vf )$ but undergoes the transformation
\beq
{\cal A} \left( g e^{it_3 \vf }\right)
= {\cal A} (g) ~+~ \delta \int {d^3x\over v} \vf
\label{9}
\eeq
Since ${\cal A}$ changes by a gradient term, its curl
$\Omega$ is invariant under this transformation. Thus
while ${\cal A}$ depends on all three $SU(2)$ variables,
$\Omega$ can be
written in terms of $SU(2)/U(1)$ variables
only.

Consider the left
translations of
$g$ given by
$\delta g = i t_a \xi^a g$, where $\xi^a$ denote infinitesimal changes of
the group parameters. These transformations may be represented by the
vector field
\beq
V (\xi )= - \int d^3x~ \xi^a (E^{-1})^i_a {\delta \over \delta
\theta^i}\label{10}
\eeq
For this transformation to be canonically generated, we must have
$\delta \left( V(\xi )\rfloor \Omega \right) =0$.
where $V\rfloor \Omega$ denotes the interior contraction of
$V$ with $\Omega$. In the present case, the topology of the phase
space ${\cal P}$ is such that this requirement is equivalent
to $ V(\xi )\rfloor \Omega =-\delta G$.
$G$ is the canonical
generator of the transformation. Explicitly we find
\beqar
V(\xi )\rfloor \Omega &=& -\delta \phi (\xi )\nonumber\\
\phi (\xi )&=& \int {d^3x \over v}~ \phi_a \xi^a \nonumber\\
\phi_a &\equiv& \Tr ( g \sigma_3 g^{-1} t_a )
\label{11}
\eeqar
Thus the left translations of $g$ can indeed be canonically generated
and the generating function is $\phi_a$. Notice that
$\phi_a$ are invariant under $g \rightarrow g e^{it_3 \vf }$ and so
are defined entirely in terms of the phase space variables.
Indeed $\phi_a$ may be taken as giving coordinates on $S^2 =
SU(2)/U(1)$ since $\phi_a \phi_a =1$. Since $\phi_a$ give a
coordinatization of the phase space, all physical observables may be
taken as functions of $\phi_a$.

A general parametrization of $g$, which is valid everywhere
except near one of the poles, is
\beq
g= {1\over \sqrt{1+\bz z}}\left( \matrix{1 & z\cr -\bz &1\cr}\right)
~\left(\matrix{e^{i{\chi \over 2}}&0\cr 0&e^{-i{\chi \over 2}}\cr}\right)
\label{12}
\eeq
The variables $\phi_a$ in this parametrization are
\beqar
\phi_1 = - {z +\bz \over 1+\bz z},~~~~&&~~~~~~~~~~
\phi_2 = i {\bz -z \over 1+\bz z}\nonumber\\
\phi_3 &=& {1-\bz z\over 1+\bz z}\label{13}
\eeqar
In these coordinates, the symplectic two-form $\Omega$
is given as
\beq
\Omega =
2i \int {d^3x \over v}
{\delta \bz \wedge \delta z\over (1+\bz z)^2}
\label{14}
\eeq
which leads to the definition of Poisson brackets as
\beq
\{ F , G \} = i{v\over 2}\int d^3x ~(1+\bz z)^2 \left( {\delta F
\over\delta \bz}
{\delta G \over \delta z}- {\delta F \over \delta z} {\delta G
\over \delta \bz} \right)\label{15}
\eeq
This immediately leads to the relations
\beq
\{ \phi_a (x) , \phi_b (y) \} = -
~\epsilon_{abc}\phi_c(x) ~\delta^{(3)} (x-y)
\label{16}
\eeq
This shows that we can identify the spin variables as
$S_a (x) = -\phi_a (x)$. In the quantum theory,
by the usual correspondence rules,
the commutation rules for these can be
written as
\beq
[ S_a (x) , S_b (y) ] = i  ~\epsilon_{abc}S_c(x)
~\delta^{(3)}(x-y)
\label{17}
\eeq
in agreement with the continuum version of the rules
(\ref{1}). We see that the action (\ref{4}) does indeed lead to
the correct commutation rules.
As mentioned before, all
physical observables are functions of $\phi_a$ or $S_a(x)$.
In particular $H$ is a function of $S_a$ and their gradients.

Notice that the expression for $\Omega$ in (\ref{14})
is the area element on the two-dimensional sphere.
The corresponding potential ${\cal A}$ is thus the potential for
a monopole with its attendant Dirac string singularities.
With $\phi^a$ providing a coordinatization of a two-sphere,
equations (\ref{14}-\ref{17}) show that we have the usual description of
the ferromagnet. Thus our considerations, as mentioned in the
introduction, can indeed be applied to physical situations such as the
quantum Hall ferromagnet.

We now turn to the momentum density $J_i$ which is expected to generate
the coordinate transformation $x^i \rightarrow x^i +a^i (x)$, or
$g \rightarrow g + a^i \partial_i g$. We may write the corresponding
vector field as
\beq
V(a) = \int d^3x~ a^i~ {\del \theta^k\over \del x^i}~{\delta \over \delta
\theta^k (x)} \label{18}
\eeq
Contracting this with $\Omega$ we get,
\beqar
V(a) \rfloor \Omega &=& - \delta \int {d^3x \over v}~ a^i \left[
-i~ \Tr (\sigma_3 g^{-1} \del_i g ) \right] \nonumber\\
&&\hskip .5in +i \int {d^3x \over v}~ (\del_i a^i )~
\Tr (\sigma_3 g^{-1} \delta g) \label{19}
\eeqar
We see that $V(a) \rfloor \Omega$ cannot be taken as
the variation of some functional because of the second
term on the right hand side in (\ref{19}).
In particular, $\delta \bigl( V(a) \rfloor \Omega \bigr)$ is not zero.
This proves that
the transformation $g \rightarrow g + a^i \del_i g$
{\it cannot} be generated as a canonical transformation.
Therefore, one cannot define a momentum density for the
ferromagnet. However, the same equation shows that
transformations for which $\del_i a^i =0$ can be canonically generated
since the second term in (\ref{19}) vanishes and the corresponding
generator is given by
\beq
J (a)= -i \int {d^3x \over v}~ a^i ~\Tr (\sigma_3 g^{-1} \del_i g)
\label{20}
\eeq
The condition $\del_i a^i=0$ corresponds to
the volume-preserving transformations. Thus, our conclusion is that only
volume preserving transformations can be canonically implemented
for the ferromagnet. The total momentum and the
orbital angular momentum
are included in this set.

As mentioned in the introduction,
this peculiarity of the momentum density for
the ferromagnet has been noticed and analyzed in a
different way before. The canonical two-form
$\Omega$ is the area element on the two-sphere
$SU(2)/U(1)$ which is the target of the mappings
$\phi_a :{\bf R}^3 \rightarrow S^2$ which constitute
the phase space. As such it has the same form as the
magnetic field of a Dirac monopole on a sphere
of fixed radius.

Using the explicit form of the $\phi_{a}$ as given in (\ref{13}),
it is easy to show that $\Omega $, the symplectic two-form
given in (\ref{14}), may be written as
\beq
\Omega = {1\over 2}~\int {{d^3x}\over{v}}\  \epsilon_{abc} \ \phi_{a} \
\delta
\phi_{b} \wedge
\delta \phi_{c} .
\eeq
Recalling that the $\phi_{a}$ provide a coordinatization of the phase space
$S^{2}$, this shows that $\Omega $ is indeed an area element on $S^{2}$. Consequently,
the symplectic potential $\cal{A}$ is nothing but the potential due to a magnetic
monopole located at the center of $S^{2}$.

The evaluation of a potential for such a field
has the well known Dirac string singularity.
Our potential ${\cal A}$ does not show this because
it is defined on a larger space; this is
evident from its
lack of invariance under $g \rightarrow g e^{it_3 \vf}$.
The difficulty in defining the momentum density is
related to this. One can define a momentum
density by enlarging the phase space. This is
best seen by comparison with the Lagrangian for
fluid motion, to which we turn now.

\section{Fluid mechanics}

In three spatial dimensions, one can parametrize the local
velocity of a fluid as
\beq
v_i = \del_i \chi ~+~ {\alpha} \del_i \beta \label{21}
\eeq
This is the standard Clebsch parametrization and can be used
to write a Lagrangian for fluid mechanics \cite{fluid, fluid2}. With zero
viscosity, the appropriate action is
\beq
{\cal S}= \int dt~d^3x~ \left[ \rho \left({\dot \chi}
+\alpha {\dot \beta}
\right) ~-~ {1\over 2} \rho \left( \del \chi +
\alpha \del \beta \right)^2
\right] \label{22}
\eeq
The first term identifies the canonically conjugate pairs
of variables $(\rho ,\chi ),~( \rho \alpha, \beta )$.
It is easily verified that the equations of motion
as the Euler equation and the equation of continuity.
It has been noted for sometime that one can write the Clebsch
parametrization as \cite{fluid2}
\beq
v_i = -i \Tr ( \sigma_3 g^{-1} \del_i g)\label{23}
\eeq
where $g$ is an element of $SU(2)$. In fact with the
parametrization (\ref{12}) one can verify this explicitly
with the identifications $\chi \leftrightarrow \chi$,
$z = r \exp (i\beta )$, $\alpha = 2r^2 /(1+r^2)$.
Thus we may write the action for fluids as
\beq
{\cal S} = -i \int d^3x~ \rho \Tr (\sigma_3 g^{-1} {\dot g}
) ~-~ \int dt ~H \label{24}
\eeq
There is a subtle specialization in this replacement
since the variables in the Clebsch parametrization
(\ref{21}) are not necessarily compact, but are replaced
by the $SU(2)$ variables which are compact.
In the quantum theory, this leads to quantization
of $\int \rho$ and the vorticity.
We are therefore considering fluids with an underlying
particle structure and which have quantized vorticity.
(One may even make a case that this is a better
modelling of actual fluids.)

The phase space ${\cal P}_F$ for the fluid
is thus the set of all maps
of the form ${\bf R}^3\rightarrow {\bf R}^4$
where $\rho \in {\bf R}_+$ and $g \in SU(2)=S^3$
form the four-dimensional space ${\bf R}^4$.
The canonical one-form and two-form for the
action (\ref{24}) are easily found to be
\beqar
{\cal A} &=& -i \int d^3x~ \rho ~\Tr ( \sigma_3 g^{-1} \delta g)
\nonumber\\
\Omega &=& -i \int d^3x~ \left[ \delta \rho ~\Tr (\sigma_3 g^{-1}
\delta g) ~-~ \rho~ \Tr (\sigma_3 g^{-1} \delta g ~g^{-1} \delta g)
\right]
\label{25}
\eeqar
The contraction of $V(\xi )$ of
(\ref{10}) with this $\Omega$ leads to
$V(\xi )\rfloor \Omega = -\delta \int \rho \phi_a
\xi^a$. Thus the generator of left translations of
$g$ is now given by $\rho \phi_a$. The variable
$S_a (x) = -\rho \phi_a$ obeys the Poisson bracket
relations
\beq
\{ S_a (x) , S_b (y) \} = \ep_{abc} ~S_c (x) \delta^{(3)}
(x-y) \label{26}
\eeq
The transformation $g \rightarrow g + a^i \del_ig$ has to be augmented
by the transformation of $\rho$. By considering
the contraction of $V(a)$ with $\Omega$, we can see that
the transformation which can be implemented
canonically is given by
\beq
V(a) + {\tilde V(a)}=
V(a) ~+~ \int d^3x~ \del_i ( \rho a^i ) {\delta \over \delta \rho}
\label{27}
\eeq
For this vector field, we find
\beq
(V(a) + {\tilde V(a)})\rfloor \Omega
= -\delta \int d^3x~ a^i~\left[ -i ~\rho~ \Tr ( \sigma_3 g^{-1}
\del_i g ) \right]\label{28}
\eeq
which identifies the momentum density as
\beq
J_i = -i ~\rho~ \Tr ( \sigma_3 g^{-1} \del_i g )
\label{29}
\eeq
The complete set of Poisson bracket realtions can then be worked out as
\beq
\begin{array}{r c l}
\{ S(\xi ) , S(\xi') \} &=& S(\xi \times \xi')\\
\{ \rho (f) , S(\xi )\}&=& 0\\
\{ \rho (f) , \rho (h) \}&=& 0\\
\{ \rho (f) , g (x) \} &=& -i g(x) ~t_3~f(x)  \\
\{ J(a), \rho (f) \}&=& \rho (a\cdot \del f )\\
\{ J(a), g (x) \}&=& - a\cdot \del g (x) \\
\{ J(a) , J(b) \} &=&  J( a\cdot \del b - b\cdot \del a )\\
\end{array}\label{30}
\eeq
$(\xi \times \xi' )^a =\ep^{abc} \xi^b \xi'^c$.

A brief aside on the quantum theory will be useful
at this stage. In the quantum theory, the above
bracket relations become commutation rules for the corresponding
operators. Consider the unitary transformation given
by $U = \exp (- 2\pi i C)$, $ C= \int \rho$. Since this corresponds to
$f = 2\pi$, we see that $S(\xi )$ and $J(a)$ are
invariant under $U$. The only variable which changes
is $g$, with $U^\dagger g U= g \exp (i\pi \sigma_3 )
=-g$. Since all the physical variables involve
even numbers of $g$'s, we see that $U$ can be taken as the identity
operator. Thus the angular nature of the variable $\chi$
in $g$ leads to the quantization condition $\int \rho =N$
for some integer $N$. This is part of the quantization
condition alluded to earlier.

It is clear from the choice of fluid variables and
the bracket relations (\ref{30}) that,
as far the canonical structure is concerned,
the ferromagnet can be considered as an incompressible
fluid for which $\rho$ is a constant, to be set to
$1/v$. Thus a reduction of the phase space ${\cal P}_F$
of fluid mechanics by the constraint
$\rho \approx 1/v$ will lead to
the ferromagnet. This shows why
only volume-preserving transformations
$x^i \rightarrow x^i +a^i$ can be canonically implemented,
since the density has been fixed in the reduction of the
phase space.

The bracket relations (\ref{30}) show that $\rho$ is the generator
of transformations of the form $g\rightarrow g e^{it_3 \vf}$.
This may be regarded as a `gauge' transformation,
to be factored out to go to the physical
phase space ${\cal P}$ for the
ferromagnet. In the canonical setting,
factoring out the gauge transformations
$g \rightarrow g e^{it_3 \vf}$ is achieved by setting
$\rho \approx 1/v$ and also imposing a
conjugate, gauge-fixing constraint and then defining
Dirac brackets.
The variables $S_a$ are not affected since they
are invariant under the gauge transformations (i.e.,
$\rho$ commutes with them) and so they
project down unaltered in form to
${\cal P}$.
$J(a)$ are not `gauge-invariant',
and so the form of $J(a)$ will be changed by the projection
to the phase space ${\cal P}$.
One has to choose a gauge in which the projected $J(a)$ are
defined. There will be no unique expression
for the projected $J(a)$. Another related
difficulty is that
any gauge-fixing has singularities \cite{haldane}. This is most easily
seen by noting that the curl of $J_i$ is well defined even after we set
$\rho$ to a constant,
\beq
dJ = \half (\del_i J_j - \del_j J_i) ~dx^i\wedge dx^j
= i \Tr (\sigma_3 g^{-1} dg \wedge g^{-1}dg)
\label{31}
\eeq
The right hand side is invariant under
$g \rightarrow ge^{it_3 \vf}$; it is
the expression for the field of
a monopole. If we can solve this equation for
$J_i$ expressing it entirely in terms of
$SU(2)/U(1)$ variables, there is no
difficulty with defining everything on ${\cal P}$.
But solving (\ref{31}) requires a gauge choice for
$J_i$ and will involve the Dirac string.
There is no gauge choice which can avoid this
singularity.
A patchwise, nonsingular solution is
possible if we can move the string around by
gauge transformations, which is effectively what is
achieved by keeping the $(\rho ,\chi )$-degrees of
freedom.

In conclusion, for the ferromagnet,
a canonical generator of coordinate
transformations can be defined only if we restrict to
volume-preserving transformations. Secondly,
the ferromagnet may be regarded as an incompressible
fluid. This analogy allows one to define
the generator of arbitrary
coordinate transformations (or momentum density)
in an enlarged phase space ${\cal P}_F$. Any choice of constraints
which allows us to go back to the ferromagnet
will have singularities and lead to singular
and nonunique expressions
for the momentum density.

\section{Topology of the phase space}

It is well known that the topology of the phase space
can lead to quantization conditions and to
the existence of multivalued wave functions, etc \cite{bal}.
We will now consider how such possibilities could arise
for the ferromagnet.

We start by considering the quantization of a
general classical
theory
formulated in the following manner.
We first consider arbitrary
complex valued functions on the full phase space
$\P$ of
the theory which are also square integrable
over the full phase space; these are the prequantum wavefunctions.
(Strictly speaking, the prequantum wavefunctions are sections of a line
bundle on the phase space, the curvature of the bundle being given by
the symplectic two-form.)
One can then obtain an explicit representation of
classical observables as operators on such functions, with the commutator
algebra reproducing the Poisson bracket relations. Then one can impose a
condition, the so-called polarization condition, which restricts the
prequantum wavefunctions to half of  the phase space
degrees of freedom. This restricted set defines the
Hilbert space of the quantum theory.
What we have outlined is the geometric quantization
of the theory. In this approach,
since the prequantum wavefunctions are
functions on the phase space, double-valuedness or
multivaluedness
requires that there be noncontractible curves
on the phase space. In other words, we need
$\Pi_1(\P )\neq 0$.

For a $d$-dimensional ferromagnet, the phase space
${\cal P}$ consists of maps from space ${\bf R}^d$
to $SU(2)/U(1)$. We will be interested in the cases
$d=2,~3$. In the ground state of the ferromagnet we have
a fixed magnetization, $\la \phi_a (x) \ra = \delta_{a3}$.
For excitations in a large enough sample, we can assume
that we have the condition $\phi_a (x) = \delta_{a3}$
at the boundary of space. the set of maps ${\bf R}^3 \rightarrow
S^2$ with this boundary condition is equivalent to
the set of maps $S^d \rightarrow S^2$. Thus
\beq
{\cal P} = \{ {\rm set ~ of ~ all ~maps:} S^d \rightarrow
S^2\}\label{31a}
\eeq

We now turn to the connectivity of this space.
The set of maps $S^d \rightarrow S^2$ can be classified
by the homotopy group $\Pi_d (S^2) = {\bf Z}$ for $d=2, 3$.
Thus the phase space ${\cal P}$ consists of a series of
mutually disconnected
components, each connected component being labelled by
the winding number
corresponding to the elements of $\Pi_d (S^2)$.
As noted before, we can use $\phi_a$, $a=1, 2, 3$,
as coordinates for the
target space two-sphere, with $\phi_a \phi_a =1$.
A point on
${\cal P}$ is thus given by a map
$\phi_a (x): \vx \rightarrow S^2$.
For most of
what follows, we can restrict attention to one of the
connected components of ${\cal P}$,
say the topologically trivial one ${\cal P}_0$
which admits the (ground state) configuration
$\phi_a (x) = \delta_{a3}$.

Consider now a closed curve in ${\cal P}_0$ which starts with
$\phi_a (x ) =\delta_{a3}$  goes through a sequence of
configurations, keeping $\phi_a \rightarrow \delta_{a3}$
as $|\vx | \rightarrow \infty$, and finally returning to
$\phi_a (x) = \delta_{a3}$. Parametrizing the curve
by $\lambda$, $0\leq \lambda \leq 1$, we see that this loop
in ${\cal P}_0$ is described by $\phi_a (x, \lambda )$ with
$\phi_a \rightarrow \delta_{a3}$ at the boundary of the
region $(x, \lambda)$. It is thus a map from $S^{d+1}$ to
the target space $S^2$. The homotopy classes of such
closed loops give $\Pi_1 ({\cal P}_0)$; in our case, this is
given by
$\Pi_{d+1}( S^2)$ from the argument given here.
Now $\Pi_3 (S^2) ={\bf Z} $ and $\Pi_4 (S^2)= {\bf Z}_2$,
so that we can expect multivalued prequantum
wavefunctions for two spatial dimensions and double-valued
wavefunctions for three spatial dimensions. One can expect
this to lead to arbitrary statistics or anyons
for certain excitations in two spatial dimensions
and fermionic excitations (as well as bosons)
in three dimensions.
(The topology associated with $\Pi_3 (S^2) ={\bf Z} $
has been used to construct anyons before,
but there are differences in our case, since our discussion involves the phase space
and not the configuration space.
The use of
$\Pi_4 (S^2) ={\bf Z}_2 $ is new, as far as we know, although
$\Pi_4 (S^3)= {\bf Z}_2$ has been used for the configuration
space of Skyrmions.

The techniques for
incorporating topological features of the configuration
space into the quantum theory
are standard by now. But there are some unusual
elements in the analysis when it is the phase space which
shows such
nontrivial connectivity. We will therefore
discuss how this can be done,
taking the two-dimensional case as an illustrative example.
In this case, $\Pi_1 ({\cal P}_0 ) = \Pi_3 (S^2) ={\bf Z}$.
This shows that there is a vector potential or one-form on the phase
space which is closed or has zero curl, but which cannot be written as
a phase space gradient or $\delta $ of some functional.
This flat potential can lead to inequivalent
representations of
the basic commutation algebra of operators.
We will now carry out the quantization incorporating this
feature.

\section{The flat potential for two spatial dimensions}

We start with the construction of the flat potential
in the case of two spatial dimensions. Since
the relevant topology is
$\Pi_3 (S^2) ={\bf Z}$, the associated
invariant is the Hopf invariant \cite{wilczek}.
We first construct it on the classical phase space
${\cal P}$.
This consists of maps from $S^2$ to $S^2$ and falls into mutually
disjoint sectors labelled by the winding number
\beqar
Q &\equiv& \int d^2x~ J_0 (x) \nonumber\\
&=& {1\over 8\pi}\int d^2x~ \epsilon^{ij}\epsilon_{abc} \phi_a \del_i
\phi_b
\del_j
\phi_c \nonumber\\
&=& {i\over 4\pi} \int \Tr (\sigma_3 g^{-1} dg ~g^{-1} dg )
\label{32}
\eeqar
The associated current is given by
\beqar
J^{\alpha} &=& {1\over 8\pi} \epsilon^{\alpha\mu\nu}\epsilon_{abc}
\phi_a \del_\mu \phi_b \del_\nu \phi_c\nonumber\\
&=& {i\over 4\pi} \epsilon^{\alpha \mu\nu}
\Tr (\sigma_3 g^{-1}\del_\mu g
~g^{-1}\del_\nu g)\label{33}
\eeqar
The Hopf invariant can now be defined as
\beq
S_{Hopf} = \int A_\mu J^\mu \label{34}
\eeq
where the gauge
potential $A_\mu$ is defined by the Chern-Simons equation
\beq
\del_\mu A_\nu -\del_\nu A_\mu = \epsilon_{\mu\nu\alpha} J^\alpha
\label{35}
\eeq
For our purpose, it is best to solve this in the gauge where
$\nabla_i A_i =0$ which works out as
\beqar
A_0 (x) &=& - \epsilon_{ik} \nabla_k \int  d^2y~ G(x,y)
J_i (y) \nonumber\\
A_i (x)&=& ~\epsilon_{ik} \nabla_k \int  d^2y~ G(x,y) J_0 (y)
\label{36}
\eeqar
where the $G(x,y)$ is the Green's function
\beq
G(x,y) = \int {d^2k \over (2\pi )^2} {e^{ik\cdot (x-y) }
\over k^2}\label{37}
\eeq
We then get the Hopf invariant as
\beqar
S_{Hopf} &=&  \epsilon_{ij} \int J_0(x) \nabla_i G(x,y)
J_j(y)\nonumber\\
&=& {i\over 2\pi} \int d^2xd^2yd\lambda~
J_0(x) G(x,y) \Tr [ g^{-1}\nabla_i g ,
\sigma_3] g^{-1} \del_\lambda g \nonumber\\
&=& {1\over 2\pi} \int d^2xd^2yd\lambda~
J_0 (x) \nabla_i G(x,y) \nabla_i \phi_a \theta^a\label{38}
\eeqar
where $\del_\lambda g g^{-1}= -i\theta^a t_a$, $t_a={\sigma_a/2}$. Notice
that this is in the form of the integral along the $\lambda$-direction of
a potential
\beq
\C (\xi ) =\int d^2y~\C_a \xi_a
= {1\over 2\pi} \int d^2x d^2y~ J_0 (x) \nabla_i G(x,y)
\nabla_i\phi_a \xi^a\label{39}
\eeq
(Here $\lambda$ denotes the parameter of a curve in the
phase space, along which we are integrating. In writing
the Hopf invariant as a term in the action,
as is done for some theories,
this parameter can be taken as time. We have in fact used this
for notational simplicity
in writing the currents in (\ref{33}-\ref{36}).)

Let $L_a$ denote the left-variation of
$g$ given by $L_a(x) g(y)= t_a \delta^{(2)}(x-y) g$.
Since $\C$ itself arises from the left-variation of the Hopf term, it is
easily verified that it obeys the flatness condition
\beq
\delta_\xi \C(\xi') - \delta_{\xi'} \C_\xi
+\C (\xi \times \xi' )=0
\label{40}
\eeq
where $\delta_\xi = \int d^2x~ \xi^a (x) L_a (x)$
and $(\xi \times \xi')^a=\epsilon^{abc} \xi^b \xi'^c$.

We now discuss how this classically available flat potential can
be incorporated in the quantum theory. This is not easily done, because
$\C$ involves the phase space coordinates and so becomes an operator
in any representation in general. (This does not happen
in $\sigma$-models where $SU(2)/U(1)$ is just the coordinate
part of the phase space; in that case, all quantities appearing in the
expression for $\C$ commute among themselves and $\C$ is
realized as a $c$-number function in the Schr\"odinger
representation.)
There are  two ways to take account of $\C$ in our theory.
The first involves the use of coherent states on the coset
$SU(2)/U(1)$ so that the difference
between the prequantum wavefunctions and the true wavefunctions is
essentially a holomorphicity requirement. The second approach will
involve writing an operator version of the potential (\ref{39}).
We will discuss both these approaches briefly.

In the first approach, the wavefunctions are taken to be functionals of
$g$, of the form $\Psi [g]$. On these, we can define right translations
$R_a$
and left translations $L_a$.
As functional differential operators,
these are given by
\beqar
L_a &=& i \left(E^{-1}\right)^i_a {\delta \over \delta
\theta^i}\nonumber\\
R_a &=& i \left( {\tilde E}^{-1} \right)^i_a {\delta
\over \delta
\theta^i}\label{41}
\eeqar
In these expressions, $\theta^i$ denote
the group parameters. The action of these translations
on $g$ is given by $L_a g =t_a g$, $R_a g =g t_a$.

Corresponding to the
symplectic potential (\ref{5}), namely,
${\cal A} = -(i/v) \int d^2x~\Tr (\sigma_3 g^{-1}\delta g)$,
treated as a vector
potential, we can define covariant
derivatives on the phase space as
${\cal R}_\pm = R_\pm$, ${\cal R}_3 = R_3 -(1/v )$.
The condition that the phase space is really $SU(2)/U(1)$
and not $SU(2)$ is expressed by
${\tilde R}_3 \Psi [g]=0$. The polarization or
holomorphicity condition can be
taken as ${\tilde R}_+ \Psi [g]=0$.

The prequantum operators corresponding to $\phi_a$
(or $S_a = -\phi_a /v$)
will be given as functional differential operators acting
on functions of $g$. From the general rules they
are seen to be given in terms of the left translations operators
$L_a$ as
\beq
{\hat \phi}_a \Psi [g] = -v~ L_a \Psi [g]
=-iv \left( E^{-1} \right)
^i_a {\delta \over \delta \vf^i} \Psi [g]\label{43}
\eeq
where $\delta g g^{-1} = it_a E^a_i d\vf^i$, $L_a g = t_a g$.
The prequantum operators can be used as the true quantum operators only
if the polarization condition commutes with them. This is the case for
us, since left and right translations commute in general. Thus we may use
the differential operators (\ref{43}) as the quantum version of
$\phi_a$ for the coherent state representation for the $\Psi$'s defined
by the conditions
\beqar
R_3 \Psi [g]&=&0\nonumber\\
R_+ \Psi [g]&=&0\label{44}
\eeqar

We can now obtain a different representation of the commutation rules
(\ref{16}) by adding a term proportional to the
flat potential $\C$ to
the left translation operators, i.e.,
\beq
{\hat \phi}_a = -v \left( L_a + k \A_a\right) \label{45}
\eeq
where $k$ is a parameter
which labels the statistics of the particles
involved.
Clearly, since $L_a$ is the left translation
operator, the commutation rules (\ref{16}) are
satisfied by virtue of the flatness condition (\ref{40}).
(Notice that we can write $\C_a =-i L_a S_{Hopf}$.)
We can thus use this representation for various
observables, in particular, in the Hamiltonian
to construct the Schr\"odinger equation for the
wavefunctions.
The physics of the problem will clearly depend on
the choice of the representation or the parameter $k$
and amounts to a different quantization of the
theory.
However, there is a difficulty which has to do with the
polarization condition. While $R_a$ and $L_b$ commute and
so the use of $L_b$ for ${\hat \phi}_b$ preserve
the conditions (\ref{44}), this is not the case once we use the
modified representation (\ref{45}).
But this can be taken care of by using
a modified version of the right translations as well.
By taking the right variation of the Hopf invariant, we
get a potential
\beqar
{\tilde \C}_a (y) &=& {1\over 2\pi} \int d^2x~
J_0 (x) \nabla_i G(x,y) \Tr \left( g^{-1}\nabla_i
g [ \sigma_3 , t_a]\right)\nonumber\\
&=& -i R_a S_{Hopf}\label{46}
\eeqar
This is a flat potential for right translations and,
using this, we obtain a new representation for the covariant
derivatives as $ {\cal R}_a + k {\tilde \C}_a$.
It is now easy to see that
\beq
[ L_a + k \C_a , {\tilde R}_b +k {\tilde \C}_b ]=0
\label{47}
\eeq
since $\C_a \sim L_a S_{Hopf},~ {\tilde \C}_a
\sim R_a S_{Hopf}$.
Thus, instead of (\ref{44}), we can use the conditions
\beqar
\left( R_3 +k {\tilde \C}_3 -{1\over v}\right)~\Psi [g] &=&0\nonumber\\
\left( R_+ +k {\tilde \C}_+ \right)~ \Psi [g] &=&0\label{48}
\eeqar
The general solution to this is of the form
\beq
\Psi [g] =\exp\left(ik \int d\lambda \left[{\tilde \C}_+
2i\Tr (g^{-1}\del_\lambda g t_-) ~+~{\tilde \C}_3 2i \Tr
(g^{-1}\del_\lambda g t_3) \right]\right)~\Phi [g]\label{49}
\eeq
where $\Phi$ obeys the old conditions (\ref{44}).
Notice that we are integrating
the flat potentials ${\tilde \C}_a$ along
two of the group directions;
if we did this for all three,
we would simply have $e^{ikS_{Hopf}}$.
We may therefore write
\beq
\Psi [g] = \exp\left(ikS_{Hopf} - ik\Theta \right)~\Phi [g]
\label{50}
\eeq
where
\beq
\Theta =2i \int d\lambda ~{\tilde \C}_-
~\Tr (g^{-1}\del_\lambda g t_+)\label{51}
\eeq
The action of ${\hat \phi}_a$ now simplifies as
\beq
(L_a + k \C_a )\Psi = L_a e^{-ik\Theta} \Phi [g]
\label{52}
\eeq
Thus the effect of the extra term is the
phase factor
$e^{-ik\Theta}$.

We have thus obtained a coherent state approach to
the quantization where the multivaluedness due to
$\Pi_1 (\P )= {\bf Z}$ can be taken account of
explicitly by a phase factor, the formulation being
compatible with the polarization requirement
on the wavefunctions.

The second way of incorporating the
effect of $\A_a$ requires the operator version of (\ref{39}).
The operator corresponding to the topological charge density
may be defined by the commutation rule
\beq
[\phi_a (x) , J_0 (f)] = -i{v \over 4\pi} ~\epsilon^{ij}
\nabla_i f
\nabla_j \phi_a (x) \label{53}
\eeq
where $J_0 (f) = \int d^2x~ J_0 (x) f(x)$. We see easily
that the topological charge commutes with $\phi_a (x)$,
and hence with any local observable, so that it is superselected,
as expected for a  topological charge.
We can also check that the expression
\beq
J_0 = {1\over 8\pi} \epsilon^{ij} \epsilon_{abc}~ \nabla_i \phi_b~
\phi_a  ~\nabla_j \phi_c
\label{54}
\eeq
can be used as the operator expression with the ordering
indicated. If singularities involved in the commutator with
$\phi_a (x)$ are regulated in an angular symmetric way,
this does lead
to the commutation rule (\ref{53}).

Consider now the expression for $\C_a$; we write this
as
\beqar
\C_a(x) &=& {1\over 2\pi} \int d^2y ~J_0 (y) \nabla_i G_{Reg}(y,x)
\nabla_i \phi_a (x) \nonumber\\
G_{Reg} (x,y) &=& \int {d^2p\over (2\pi )^2}{ e^{ip\cdot (x-y)}
\over p^2}~e^{-p^2/M^2}\label{55}
\eeqar
As the regulator parameter $M\rightarrow \infty$ this reverts
to the expression (\ref{39}). The regularization is angular
symmetric; there is no ordering ambiguity in the expression
for $\C_a$ with this regulator.
We can now define an operator $\Lambda$ by the commutation relation
\beq
[\phi_a (x) , \Lambda ] = -iv ~\C_a (x) \label{56}
\eeq
In terms of $\Lambda$ we can construct $U= \exp (-ik \Lambda )$.
The new representation is then given by ${\hat \phi}_a
= U^\dagger \phi_a U$. Eventhough formally of the form of a unitary
operator, $U$ is, in general, not unitary for arbitrary $k$
and hence this reprsentation is inequivalent to the previous one
with $k=0$.

\section{Conclusion}

We have addressed two specific problems in this paper,
both of which are
related to topological issues in the continuum version
of the ferromagnet.
The first one has to do with the definition of the momentum density.
Only moments of the momentum desnity
corresponding to volume-preserving transformations can be defined.
A more general definition is possible, in a way free of singularities,
only by enlarging the phase space. What is interesting is that the
enlarged phase space is mathematically very closely related to
the phase space for fluid dynamics. The connection between the ferromagnet
and fluid dynamics that we have emphasized should enable one to apply
the powerful techniques developed in the latter discipline to the former and
lead, in general to a successful cross-fertilization of both fields.

The second problem deals with the fact that in two and three
dimensions
the connectivity of the phase space is nontrivial. This can lead to
nontrivial statistics for excitations. But the relevant topological
term involves all the phase space coordinates which are mutually
noncommuting. There are subtleties due to this fact,
and therefore, unlike the
case of the usual sigma models, a full quantum treatment
requires additional
considerations. We have given one way of handling this situation.

\vskip .1in\noindent
{\bf Acknowledgements}
\vskip .1in
VPN's work was upported in part by
the National Science Foundation
under grant number PHY-0244873 and
by a PSC-CUNY grant. VPN also thanks D. Schmeltzer
for a useful discussion.


\begin{thebibliography}{99}

\bibitem{haldane} F.D.M. Haldane, Phys. Rev. Lett.
{\bf 57}, 1488 (1986).

\bibitem{bishop} R. Balakrishnan and A.R. Bishop,
Phys. Rev. Lett. {\bf 55}, 537 (1985).

\bibitem{sondhi}S.L.~Sondhi {\it et.al.}, Phys. Rev. B {\bf 47}, 16419
(1993); K.~Moon {\it et.al.}, Phys. Rev. B {\bf 51}, 5138 (1995); R.~Ray,
Phys. Rev. B {\bf 60}, 14154 (1999).

\bibitem{canonical} see, for example,
V.I. Arnol'd, {\it Mathematical Methods of
Classical Mechanics} (Springer-Verlag, New York, 1989);
V. Guillemin and S. Sternberg, {\it
Symplectic Techniques in Physics}, Cambridge University Press (1990).

\bibitem{fluid}
A. Clebsch, J. Reine Angew. Math. \textbf{56}, 1 (1859);
H. Lamb, \textit{Hydrodynamics} (Cambridge University Press,
Cambridge UK, 1932) p. 248; C. C. Lin in \textit{International
School of Physics E. Fermi (XXI)}, G. Careri \textit{ed.}
(Academic Press, New York, 1963); for a recent review,
see R. Jackiw,
\textit{Lectures on Fluid Dynamics} (Springer, New York, 2002).

\bibitem{fluid2} R. Jackiw and S-Y. Pi,
Phys. Rev. {\bf D61}, 105015 (2000); R. Jackiw, V. P. Nair and S.-Y. Pi,
Phys. Rev.
\textbf{D62}, 085018 (2000).

\bibitem{bal} N. Woodhouse, {\it Geometric Quantization},
Oxford University Press (1980);
A.P.
Balachandran, G. Marmo, B.S. Skagerstam and A. Stern, {\it Classical
Topology and Quantum States}, World Scientific, Singapore (1991).

\bibitem{wilczek}F.~Wilczek, Phys.Rev.Lett. {\bf 49}, 957 (1982);
F~.Wilczek and A.~Zee, Phys.Rev.Lett. {\bf 51}, 2250 (1983);
M.~Bowick, D.~Karabali and L.C.R.~Wijewardhana, Nucl.Phys. {\bf
B271}, 417 (1986); G.~Semenoff and P.~Sodano, Nucl.Phys. 
{\bf B328}, 753 (1989);
for a recent review, see F.~Wilczek, {\it Fractional
Statistics and Anyon Superconductivity}, World Scientific, Singapore
(1990).

\end{thebibliography}
\end{document}